\documentclass[aps,twocolumn,showpacs]{revtex4}
\usepackage{dcolumn}
\usepackage{graphicx}
\usepackage{amsmath}
\usepackage{amsfonts}
\usepackage{amssymb}
\usepackage{psfrag}
\usepackage{wrapfig}
\usepackage{subfigure}
\usepackage{makeidx}
\usepackage{bm}
\usepackage{epsf}
\usepackage{multirow}
\usepackage[colorlinks,urlcolor=blue,citecolor=blue]{hyperref}

\begin{document}
\title{Spin Solitons in Spin-1 Bose-Einstein Condensates}
\author{Ling-Zheng Meng, Yan-Hong Qin, Li-Chen Zhao}\email{zhaolichen3@nwu.edu.cn}
\address{School of Physics, Northwest University, Xi'an, 710069, China}
\address{Shaanxi Key Laboratory for Theoretical Physics Frontiers, Xi'an, 710069, China}
\date{\today}
\begin{abstract}
Solitons in multi-component Bose-Einstein condensates have been paid much attention, due to the stability and wide applications of them. The exact soliton solutions are usually obtained for integrable models. In this paper, we present four families of exact spin soliton solutions for non-integrable cases in spin-1 Bose-Einstein Condensates. The whole particle density is uniform for the spin solitons, which is in sharp contrast to the previously reported solitons of integrable models. The spectrum stability analysis and numerical simulation indicate the spin solitons can exist stably. The spin density redistribution happens during the collision process, which depends on the relative phase and relative velocity between spin solitons. The non-integrable properties of the systems can bring spin solitons experience weak amplitude and location oscillations after collision. These stable spin soliton excitations could be used to study the negative inertial mass of solitons, the dynamics of soliton-impurity systems, and the spin dynamics in  Bose-Einstein condensates.
\end{abstract}
\maketitle

\section{Introduction}
Multi-component Bose-Einstein condensate (BEC) provides a good platform to study dynamics of vector solitons \cite{Kevrekidis,Bersano}. Many different vector solitons have been obtained in the two-component coupled BEC systems, such as bright-bright soliton \cite{Zhang}, bright-dark soliton \cite{Nistazakis}, dark-bright soliton \cite{XBO,BEC}, and dark-dark soliton \cite{Ebd}. Those vector solitons mainly refer to particle density solitons, since the superposition of their density also admits localized wave profiles.
Recently, it was shown that it is possible to find dark-anti-dark soliton in BEC with unequal intra- and inter-species interaction strength \cite{Qu,Qu2,Danaila,Gallem}. If the superposition of the dark soliton and anti-dark soliton in the two components is uniform density, the dark-anti-dark soliton would become the so-called ``magnetic soliton"  \cite{Qu,Qu2}. Magnetic soliton is a special spin soliton for which the superposition density of components is uniform, and the spin soliton possesses a spin-balanced background. These recent developments on matter wave solitons suggest that there could be more exotic soliton profiles, which are quite different from the well-known solitons in BECs with equal intra-species and inter-species interaction strengthes   (corresponding to the integrable Manakov model \cite{Mana}).

Multi-component BEC can be seen as  pseudo-spin systems with different spin values. For example, two-component BEC can be seen as a spin-$1/2$ system \cite{SOC1,SOC2,SOC3}. Recently, spin solitons were derived in a spin-$1/2$ BEC system with different intra-species and inter-species interaction strength \cite{zhaoliu}. Spin soliton is distinctive from the general bright-dark soliton or dark-bright soliton of Manakov model \cite{Lakshman,Kanna,Lingdnls,Feng,DSWang,Hamner,Middelkamp,Yan,LCZhao,Laksh,Zhao2,Qin},  since the total mass density for spin soliton is uniform.  The spin solitons admit spin-imbalanced background,  distinguished from the ``magnetic soliton"  \cite{Qu,Qu2}.  It was demonstrated that spin soliton admit many striking properties, such as negative mass, and periodic transition between positive mass and negative mass \cite{zhaoliu}. For spin-$1$ systems, most previously reported solitons are obtained in the integrable cases (mainly including Manakov model \cite{Lakshman,Kanna,Lingdnls,Feng,DSWang} and the coupled model with spin-dependent interactions \cite{Ho,Ohmi,Ieda,Li,WZhang,Uchiyama,Kivshar,Ueda}), but spin soliton is still absent. Based on the spin soliton or magnetic solitons in spin-$1/2$ systems \cite{Qu,Qu2,zhaoliu}, we expect that the spin soliton in spin-$1$ systems could be obtained with breaking the integrable conditions.

 In this paper, we present four families of exact spin soliton solutions in spin-$1$ BEC, with proper constrained conditions on intra- and inter-species interaction strength. The constrain conditions are different from Manakov model \cite{Lakshman,Kanna,Lingdnls,Feng,DSWang} and the other integrable model with spin-dependent interactions and spin-exchange effects \cite{Ho,Ohmi,Ieda,Li,WZhang,Uchiyama,Kivshar,Ueda}. The excitation conditions for the four families spin solitons are clarified. We further perform the spectrum stability analysis on them. The results indicate that these spin solitons are stable against noises. Numerical simulations also support their evolution stability against small deviations from the constrained conditions on nonlinear interactions between atoms. The collision between spin solitons can be both elastic and inelastic, which depends on the relative phase and relative velocity between them. The spin density redistribution happens during the collision process. Spin soliton can experience weak amplitude and location oscillations after the collision. These characters are induced by the non-integrable properties of the systems. The exact spin solitons further enrich the soliton families in BECs, and provide a good benchmark for variational method and numerical simulation studies on multi-component condensate systems.

 The paper is organized as follows. In Sec. II, we present four families of spin solitons in the spin-$1$ BEC systems with proper constrains on nonlinear interaction strength. The existence conditions and spin density characters are described in details. Then, we perform stability analysis on these spin solitons in Sec. III. It is shown that spin solitons admit spectrum stability, and they are also stable against small nonlinear interaction strength deviations from the constraint conditions. In Sec. IV, we investigate the collision between spin solitons. It indicates that spin density redistribution happens during the collision process. The non-integrable characters can makes spin solitons experience weak amplitude and location oscillations after the collision. Finally, we make a conclusion in Sec. V.

\section{Three-component Bose-Einstein condensate system and spin soliton solutions}
Under the mean-field approximation,  the dynamics of three-component BEC systems in a quasi-one-dimensional limit can be described by the following coupled Gross-Pitaevskii equations \cite{Boris,Kevrekidis2}
\begin{eqnarray}\label{1}
 i \frac{\partial\Psi_{+}}{\partial t}&=&-\frac{1}{2}\frac{\partial^2\Psi_{+}}{\partial z^2}-(g_{+0} |\Psi_0|^2+g_{+-}|\Psi_-|^2)\Psi_+ \nonumber\\
 && - g_{++} |\Psi_+|^2 \Psi_+, \nonumber\\
 i \frac{\partial\Psi_{0}}{\partial t}&=&-\frac{1}{2}\frac{\partial^2\Psi_{0}}{\partial z^2}-(g_{0+} |\Psi_+|^2+g_{0-}|\Psi_-|^2)\Psi_0  \nonumber\\
 && - g_{00} |\Psi_0|^2 \Psi_0, \nonumber\\
 i \frac{\partial\Psi_{-}}{\partial t}&=&-\frac{1}{2}\frac{\partial^2\Psi_{-}}{\partial z^2}-(g_{-+} |\Psi_+|^2+g_{-0} |\Psi_0|^2)\Psi_- \nonumber\\
&&- g_{--}|\Psi_-|^2 \Psi_-,
\end{eqnarray}
where $z$ and $t$ denote the spatial coordinate and time evolution respectively. $\Psi_{\pm,0} $ refers to the three pseudo-spin components of a spin-$1$ system. The atomic mass $m$ and Planck's constant $\hbar$ are re-scaled to be $1$. The coefficients $g_{ii}$ and $g_{ij}$ ($i,j=+,0,-$) correspond to the intra- and inter-species interactions between atoms respectively. If $g_{ii}$ and $g_{ij}$ are all equal,  the dynamical equations will become Manakov model \cite{Mana}, which can be solved exactly by Darboux transformation \cite{Mat,Dok}, Hirota bilinear method \cite{Hirota}, etc.. Bright-bright and bright-dark soliton usually exist in the coupled BEC with attractive interactions \cite{Lakshman,Kanna,Lingdnls,Feng,DSWang}, while dark-bright and dark-dark soliton usually exist in the coupled BEC with repulsive interactions \cite{Hamner,Middelkamp,Yan}. The three-component BEC can be seen as a spin-$1$ system, which is usually described by the coupled model with spin-dependent interactions and spin-exchange effects \cite{Ho}.  With some certain constraint conditions, the coupled model can become an integrable system, for which many exact soliton solutions were obtained \cite{Ohmi,Ieda,Li,WZhang,Uchiyama,Kivshar,Ueda}. The soliton dynamics is usually distinctive from the ones in Manakov model \cite{Lakshman,Kanna,Lingdnls,Feng,DSWang}. It is noted that Feshbach resonance technique can be used to manipulate the scattering lengths among three hyperfine states atoms  \cite{Egorov,Widera,Thalhammer,Tojo,Mertes}. Recently, it was shown that even spatial control of the interactions between atoms can be realized \cite{Arunkumar}. These developments motivate us to consider the cases with spin-dependent interactions. The spin-exchange terms are closed, in contrast to the model in Ref.\cite{Ho,Ohmi,Ieda,Li,WZhang,Uchiyama,Kivshar,Ueda}, since the spin-exchange effects can induce particle transition between different spin components.

We note that it is possible to construct exact spin soliton solution with $g_{++}=g_{+0}=g_{0+}=g_{00}=g_{1}$, $g_{+-}=g_{-+}=g_{0-}=g_{-0}=g_{2}$ and $g_{--}=g_{3}$. For spin solitons, their total mass densities are constants \cite{zhaoliu}. For simplicity and without losing generality, we choose $|\Psi_{+}|^2+|\Psi_{0}|^2+|\Psi_{-}|^2 = 1$.
To gain insight into the existence condition of spin solitons, we focus on the nonlinear coefficient difference $g=g_{1}-g_{2}=g_{2}-g_{3}$. The spin solitons just exist with $g \neq 0$. If $g=0$, the model will become the well-known Manakov model \cite{Lakshman,Kanna,Lingdnls,Feng,DSWang}, and the spin solitons can not exist anymore. For previous integrable models, the soliton types of Manakov model usually depend on the interactions between atoms.  For example, bright solitons and dark solitons exist for nonlinear Schr\"{o}dinger equation $i\psi_{t} +\frac{1}{2}\psi_{zz}\pm |\psi|^2 \psi = 0 $, with the nonlinear coefficient takes positive and negative values, respectively. But spin soliton is different from the solitons in Manakov model, the existence of spin soliton solution depends only on the differences between nonlinear coefficients, regardless of absolute nonlinear interaction strength. Various exact soliton solutions of non-integrable models can be obtained under different parameter constraints. If $g>0$, the spin soliton mainly includes two families:  bright-bright-dark spin soliton (B-B-DSS), bright-dark-dark spin soliton (B-D-DSS); if $g<0$, there are other two families of spin solitons:  dark-dark-bright spin soliton (D-D-BSS), and dark-bright-bright spin soliton (D-B-BSS). It should be noted that B-B-DSS and D-B-BSS (B-D-DSS and D-D-BSS) are essentially different, because they admit different moving speed limitations, and be obtained with distinctive nonlinear interactions between atoms.  We emphasize that these solitons are fundamentally different from the ones reported in integrable models \cite{Lakshman,Kanna,Lingdnls,Feng,DSWang,Ohmi,Ieda,Li,WZhang,Uchiyama,Kivshar,Ueda}, although they have similar expressions.  The  fundamental differences between spin solitons and solitons in the integrable systems can be demonstrated clearly by the solitons' collision behaviors (see Sec. IV). We derive single spin soliton solutions exactly and analytically, but fail to derive the multi-spin soliton solution, due to the non-integrable properties of the systems. Moreover, It is known that the positive values of $g_{ij}$ represent the attractive interactions between atoms, and the non-zero density background  admits modulational instabilities. Therefore, we mainly discuss the case with negative values of $g_{ij}$ (corresponding to the repulsive interactions between atoms) to demonstrate our results.

\begin{figure}[htb]
\begin{center}
\includegraphics[height=60mm,width=85mm]{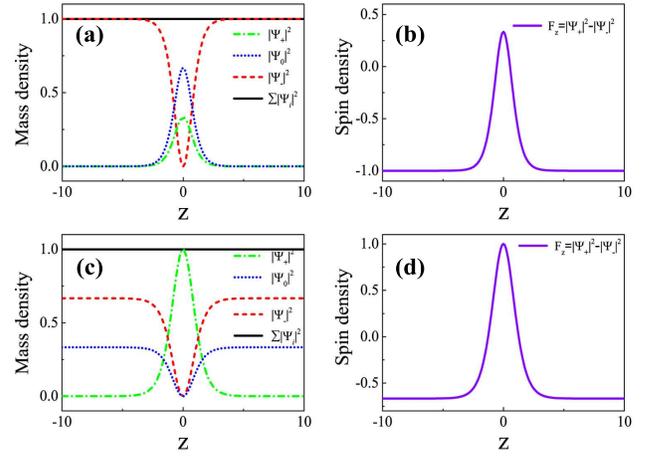}
\end{center}
\caption{The mass densities and spin densities profiles for spin solitons with $g=g_1-g_2=g_2-g_3>0$. (a) and (b) for the bright-bright-dark spin soliton (B-B-DSS); (c) and (d) for the bright-dark-dark spin soliton (B-D-DSS). The green dash-dotted line, blue dotted line and red dashed line represent $\Psi_+,\Psi_0,\Psi_-$ component respectively. The whole particle density $|\Psi_+|^2+|\Psi_0|^2+|\Psi_-|^2$ (black solid line) is uniform  for spin soliton, in contrast to the vector solitons obtained in integrable systems.  The spin density distribution admits a hump on nonzero density background. The moving speed limitation and spin density value regime are clarified clearly (see the text for details).
The parameters are $c_{1}=\sqrt{1/3}, c_{2}=\sqrt{2/3}$, $s_{1}=\sqrt{1/3}, s_{2}=\sqrt{2/3}$, and $v=0$. The nonlinear interaction coefficients are $g_{1}=-1, g_{2}=-2$, and $g_{3}=-3$.} \label{Fig1}
\end{figure}

On the one hand, we present the spin soliton solutions for the model with $g=g_{1}-g_{2}=g_{2}-g_{3}>0$,  which are B-B-DSS and B-D-DSS. The B-B-DSS solution can be written as
\begin{eqnarray}\label{2}
\Psi_{+}&=& c_{1} \sqrt{\frac{g-v^2}{g}} \textmd{sech}[\sqrt{g-v^2}(z-vt)] e^{i[vz-(v^2-\frac{g}{2}-g_{2})t]},\nonumber\\
\Psi_{0}&=&c_{2} \sqrt{\frac{g-v^2}{g}} \textmd{sech}[\sqrt{g-v^2}(z-vt)] e^{i[vz-(v^2-\frac{g}{2}-g_{2})t]}, \nonumber\\
\Psi_{-}&=& \{v-i\sqrt{g-v^2} \textmd{tanh}[\sqrt{g-v^2}(z-vt)] \} \nonumber\\
&&\frac{v+i \sqrt{g-v^2}}{g} e^{i g_{3}t}.
\end{eqnarray}
where $v$ is the velocity of spin soliton, $c_{1}$ and $c_{2}$ are used to control amplitudes of bright solitons, with $c_{1}^2+c_{2}^2=1$ holding. It is seen that the velocity $v$ must be smaller than $\sqrt{g}$, and $g$ is difference between the intra-species and inter-species interaction strength. This character is different from the vector solitons for Manakov model, for which the maximum speed is determined by the nonlinear interaction strength  \cite{Lakshman,Kanna,Lingdnls,Feng,DSWang,Ohmi,Ieda,Li,WZhang,Uchiyama,Kivshar,Ueda}. Moreover, the amplitude and width of bright soliton depends on the moving velocity, which is quite different from the bright solitons reported before. This comes from that the total particle density keeps uniform for spin solitons. As an example, the mass density and spin density profiles for a B-B-DSS with $v=0$ are shown in Fig.~\ref{Fig1} (a) and (b). The spin density background for the spin soliton is always $-1$, and the spin density maximum value can be varied in $(-1,1]$ regime by changing $c_1$ and $v$. The uniform total mass density is always absent for vector solitons obtained in integrable models \cite{Lakshman,Kanna,Lingdnls,Feng,DSWang,Ohmi,Ieda,Li,WZhang,Uchiyama,Kivshar,Ueda}, even for the non-degenerate solitons in spinor BEC reported recently \cite{LCZhao,Laksh,Zhao2,Qin}. We emphasize that the vector solitons previously obtained in integrable models still can admit soliton in spin density distribution, but the whole particle density can not be uniform. Therefore spin solitons provide possibilities to investigate separate spin current and mass current.

\begin{figure}[htb]
\begin{center}
\includegraphics[height=60mm,width=85mm]{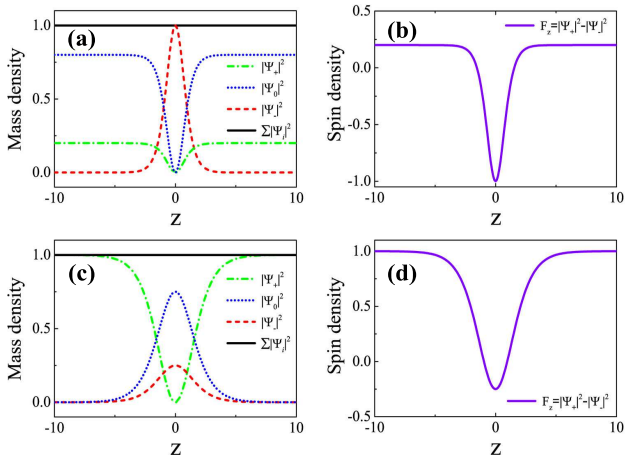}
\end{center}
\caption{The mass densities and spin densities profiles for spin solitons with $g=g_1-g_2=g_2-g_3<0$. (a) and (b) for the dark-dark-bright spin soliton (D-D-BSS); (c) and (d) for the dark-bright-bright spin soliton (D-B-BSS). The green dash-dotted line, blue dotted line and red dashed line represent $\Psi_+,\Psi_0,\Psi_-$ component respectively. The whole particle density $|\Psi_+|^2+|\Psi_0|^2+|\Psi_-|^2$ (black solid line) is uniform  for spin soliton.  The spin density distribution admits a dip on nonzero density background. The moving speed limitation and spin density value regime are clarified clearly (see the text for details).  It should be noted that B-B-DSS and D-B-BSS (B-D-DSS and D-D-BSS) are essentially different, because they admit different moving speed limitations, and can not be  related with each other by any trivial gauge transformations.
The parameters are $s_{1}=\sqrt{1/5}, s_{2}=\sqrt{4/5}, w=\frac{1}{2}$, and $v=0$, the nonlinear coefficients $g_{1}=-3, g_{2}=-2, g_{3}=-1$. }\label{Fig2}
\end{figure}

The B-D-DSS solution is given as follows
\begin{eqnarray}
\Psi_{+}&=& \sqrt{\frac{g s_{2}^2-v^2}{g s_{2}^2}} \textmd{sech}[\sqrt{g s_{2}^2-v^2}(z-vt)]\nonumber\\
 && e^{i[vz+(g s_{1}^2+\frac{1}{2}g s_{2}^2+g_{2}-v^2)t]},\nonumber\\
\Psi_{0}&=& \{v-i\sqrt{g s_{2}^2-v^2} \textmd{tanh}[\sqrt{g s_{2}^2-v^2}(z-vt)]\} \nonumber\\
&&\frac{s_{1}(v+i\sqrt{g s_{2}^2-v^2})}{g s_{2}^2} e^{i(g s_{1}^2+g_{2})t},\nonumber\\
\Psi_{-}&=& \{v-i\sqrt{g s_{2}^2-v^2} \textmd{tanh}[\sqrt{g s_{2}^2-v^2}(z-vt)]\}\nonumber\\
&& \frac{v+i\sqrt{g s_{2}^2-v^2}}{g s_{2}} e^{-i(g s_{2}^2-g_{2})t}.
\end{eqnarray}
In this case, $s_{1}$ and $s_{2}$ are introduced to vary the background amplitudes of the two dark solitons, with  $s_{1}^2+s_{2}^2=1$ holding. Parameter $v$ still denotes soliton velocity, the maximum speed is determined by the amplitude $s_2$ of the third component and the nonlinear strength difference $g$, namely $v<\sqrt{g s_{2}^2}$. This means that the maximum speed of B-D-DSS is much less than the B-B-DSS's, since $s_2$ is always smaller than $1$. As an example, we show the
mass density and spin density in Fig.~\ref{Fig1} (c) and (d). The spin density background for the spin soliton is in the $[-1,0)$ regime, and the spin density maximum value can be varied in $(0,1]$ regime by changing $s_2$ and $v$. Especially, when the spin density background value is zero, the spin soliton will admit a spin-balanced background, which is similar to the ``magnetic soliton"  \cite{Qu,Qu2}.

On the other hand, we present the spin soliton solutions for the model with $g<0$,  which are D-D-BSS and D-B-BSS.  Exact solution for D-D-BSS can be given as
\begin{eqnarray}
\Psi_{+}&=& \{v-i\sqrt{-g-v^2} \textmd{tanh}[\sqrt{-g-v^2}(z-vt)]\} \nonumber\\
&& \frac{s_{1}(v+i\sqrt{-g-v^2})}{-g} e^{i(g+g_{2})t},\nonumber\\
\Psi_{0}&=&\{v-i\sqrt{-g-v^2} \textmd{tanh}[\sqrt{-g-v^2}(z-vt)]\} \nonumber\\
&& \frac{s_{2}(v+i\sqrt{-g-v^2})}{-g}e^{i(g+g_{2})t}, \nonumber\\
\Psi_{-}&=& \sqrt{\frac{g+v^2}{g}} \textmd{sech}[\sqrt{-g-v^2}(z-vt)]\nonumber\\
&& e^{i[vz+(g_{2}-\frac{g}{2}-v^2)t]}.
\end{eqnarray}
The physical meaning of the parameters are consistent with the ones for B-D-DSS solution, and the constraint $s_{1}^2+s_{2}^2=1$ still holds. It should be noted that D-D-BSS is different from B-D-DSS, though they have similar expressions. The maximum speed for D-D-BSS depends on the interaction strength difference $g$, namely, $v<\sqrt{-g}$. This is different from the case $v<\sqrt{g s_{2}^2}$ for B-D-DSS. Therefore, the maximum moving speed of D-D-BSS can be much larger than the B-D-DSS's. One can not obtain one of them through trivial gauge transformation from the other. Moreover, the spin density background for the D-D-BSS is in the $[0,1]$ regime, and the spin density minimum value can be varied in the $[-1,1)$ regime by changing $s_{1}$ and $v$.  As an example, we show one D-D-BSS with zero velocity in Fig.~\ref{Fig2} (a) and (b). We can see that the spin density of D-D-BSS admits a ``dip", which also differs from the B-D-DSS case.

Last but not least,  the exact expression of D-B-BSS can be derived with $g<0$ as follows
\begin{eqnarray}
\Psi_{+}&=& \frac{1}{v-iw}\{v-iw\textmd{tanh}[w(z-vt)]\} e^{i(g+g_{2})t},\nonumber\\
\Psi_{0}&=& \frac{-w\sqrt{-g-(w^2+v^2)}}{\sqrt{-g}(v+iw)} \textmd{sech}[w(z-vt)] \nonumber\\ &&e^{i\{vz-[\frac{1}{2}(v^2-w^2)-g-g_{2}]t\}},\nonumber\\
\Psi_{-}&=& \frac{w}{\sqrt{-g}} \textmd{sech}[w(z-vt)] e^{i\{vz-[\frac{1}{2}(v^2-w^2)-g_{2}]t\}}.
\end{eqnarray}
The parameter $w$ denotes the width of spin soliton, $v$ is the soliton's velocity. For D-B-BSS, the limitation on moving speed is given by $\sqrt{w^2+v^2}<\sqrt{-g}$, in contrast to the above three spin solitons. We emphasize that the D-B-BSS solution can not be obtained through any trivial gauge transformation from B-B-DSS solution. The maximum speed of D-B-BSS is also different from the B-B-DSS's. Spin density background for D-B-BSS admits a fixed value $1$, and the minimum spin density value can be adjusted in the regime of $[-1,1)$.  We show the mass density and spin density of a static D-B-BSS in Fig.~\ref{Fig2} (c) and (d).

Recently, many variation methods were developed to derive the analytical soliton in BEC with different nonlinear interaction strength \cite{Katsimiga,WMLiu,Carr}. Numerical simulations were also used to investigate dynamics of soliton in non-integrable cases \cite{Tang,Wang2,wang3}. The exact analytical solutions presented above may provide a good benchmark to test the approximation results.

\section{Stability analysis of spin solitons}
\begin{figure}[htb]
\begin{center}
\includegraphics[height=55mm,width=75mm]{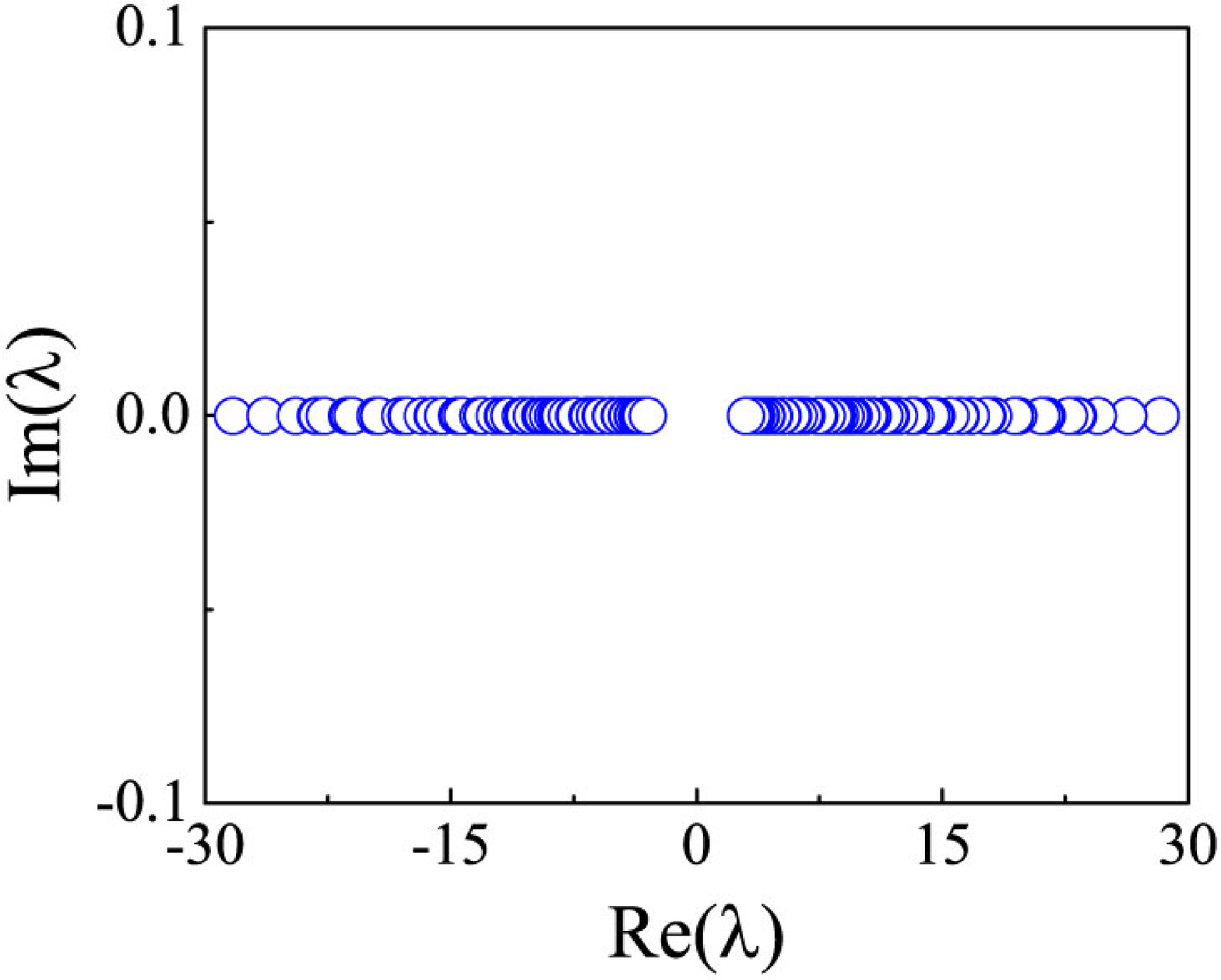}
\end{center}
\caption{The Bogoliubov-de Gennes excitation spectrum of a bright-bright-dark spin soliton. The zero imaginary parts of $\lambda$ indicates that the spin soliton admits spectral stability. The other three types of spin solitons also admit spectral stability (we do not show them anymore).  The parameters are identical with Fig. 1 (a) and (b).
}\label{Fig3}
\end{figure}

Several exact spin soliton solutions have been derived in the previous section. Then it is necessary to test the stability of them. We calculate the Bogoliubov-de Gennes excitation spectrum of a spin soliton solution  $\bm{\Psi_{sol}}=(\Psi_{+}, \Psi_{0}, \Psi_{-})^{T}$ of Eq.~(\ref{1}). We introduce perturbations on the spin soliton solution, namely, $\bm{\Psi} = \bm{\Psi_{sol}}+\epsilon(\bm{P} e^{-i \lambda t}+\bm{Q}^* e^{i \lambda^* t}) e^{i \mu t}$, where $\epsilon$ is small enough to denote weak perturbations, $\lambda$ is an eigenvalue, and $(\bm{P},\bm{Q}^*)$ define the eigenvector. Substituting $\bm{\Psi}$ into Eq.~(\ref{1}) and ignoring the high-order terms of the perturbation $\epsilon$, we can solve an eigenvalue problem for eigenvectors $(\bm{P},\bm{Q}^*)$ and eigenvalue $\lambda=\lambda_{R}+i\lambda_{I}$. The stability of spin soliton corresponds to $\lambda_{I}=0$. The calculation results show that the above four spin solitons are all stable. As an example, we show eigenvalue spectrum of a B-B-DSS in Fig.~\ref{Fig3}, where the parameters are identical with the ones in Fig.~\ref{Fig1}.  It is seen that spin solitons indeed admits spectrum stability.

\begin{figure}[htb]
\begin{center}
\includegraphics[height=25mm,width=85mm]{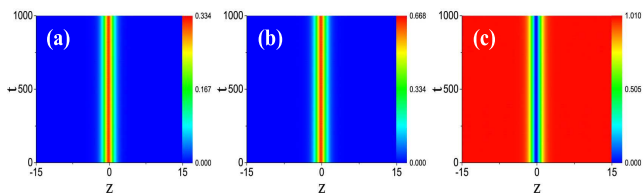}
\end{center}
\caption{The evolution of a bright-bright-dark spin soliton with the nonlinear strength admitting small deviations from the constraint conditions $g=g_{1}-g_{2}=g_{2}-g_{3}$. Panel (a), (b) and (c) denote $\Psi_{+}, \Psi_{0}$, and $\Psi_{-}$ component respectively. It is seen that the solitons' profiles in three components all kept well up to $1000$ dimensionless units.
The numerical simulations are performed with $g_{1}=-1.3, g_{2}=-2.1, g_{3}=-3.2$, and the initial soliton states are given by the exact spin soliton solution in Fig. 1 (a) and (b). }\label{Fig4}
\end{figure}

The above stability analysis is performed with $g=g_{1}-g_{2}=g_{2}-g_{3}$ condition. However, the constraint conditions on intra-species and inter-species interaction strength usually can not be satisfied perfectly by Feshbach resonance technique  \cite{Egorov,Widera,Thalhammer,Tojo,Mertes}. Therefore, we would like to further test the evolutionary stability of spin solitons with the nonlinear strength admitting small deviations from the constraint conditions. The numerical simulations indicate that the spin solitons are also stable against the small nonlinear interaction strength deviations. As an example, we show the numerical simulation results from a initial B-B-DSS in Fig.~\ref{Fig4}. The nonlinear interactions $g_{1}=-1, g_{2}=-2, g_{3}=-3$ for exact spin solitons are changed to be $g_{1}=-1.3, g_{2}=-2.1, g_{3}=-3.2$, and the initial soliton states are given by the exact spin soliton solution. It is seen that the solitons' profiles in three components all kept well over long time evolution (up to $1000$ dimensionless units). These results indicate that it is possible to excite these spin solitons in experiments, with the development of experimental techniques \cite{Bersano}.

\begin{figure}[htb]
\begin{center}
\includegraphics[height=65mm,width=85mm]{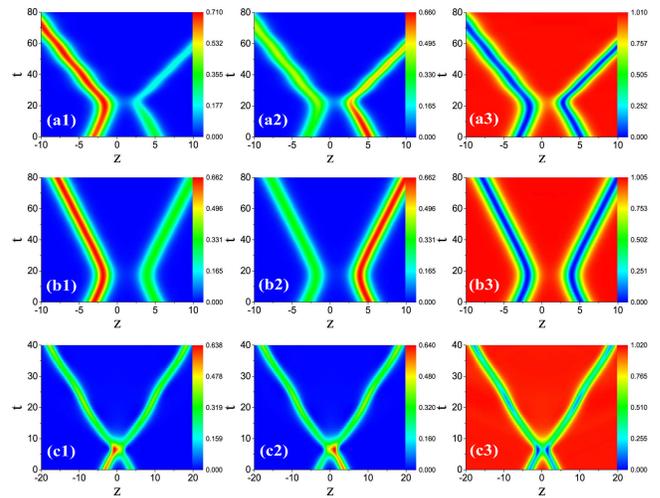}
\end{center}
\caption{The spin density redistribution induced by spin solitons' collision (their moving velocities are much lower than the speed limit). (a1-a3):  The collision between two spin solions with the velocity $v_{1}=-v_{2}=0.1$ and relative phase $\Delta\phi=\frac{\pi}{2}$;  (b1-b3): the collision between two spin solions with $v_{1}=-v_{2}=0.1$ and $\Delta\phi=0$;  (c1-c3): the collision between two spin solions with $v_{1}=-v_{2}=0.4$ and $\Delta\phi=0$. One can see that the spin density redistribution depends on the relative phase and relative velocity between spin solitons. Moreover, spin solitons can experience weak amplitude and location oscillations after collision.  This character is induced by the non-integrable properties of the system.
The other parameters are $c_{1}=\sqrt{1/3}, c_{2}=\sqrt{2/3}, dz_{1}=-3, dz_{2}=5$ and $g=1$.}\label{Fig5}
\end{figure}

\section{collisions between spin solitons}

\begin{figure}[htb]
\begin{center}
\includegraphics[height=45mm,width=85mm]{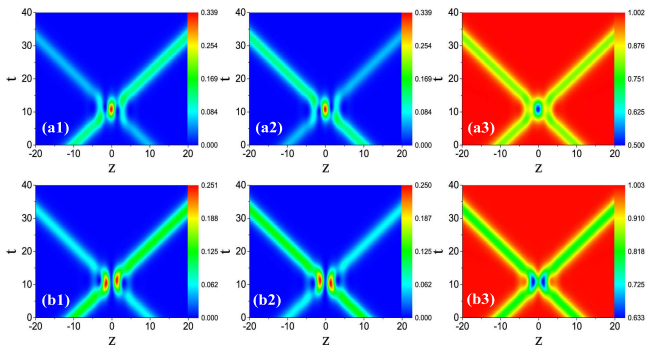}
\end{center}
\caption{The elastic collision between two spin solitons (their moving velocities are close to the speed limit). (a1-a3):  The collision between two spin solions with the velocity  $v_{1}=-v_{2}=0.9$ and relative phase $\Delta\phi=0$;  (b1-b3): the collision between two spin solions with $v_{1}=-v_{2}=0.9$  and $\Delta\phi=\pi$.  The two solitons' profiles kept well after their collision and the interference patterns emerge during the interacting process. The relative phase does not affect the elastic collision properties, but vary the interference pattern.
The other parameters are $c_{1}=\sqrt{1/3}, c_{2}=\sqrt{2/3}, z_{10}=-10, z_{20}=10$ and $g=1$.}\label{Fig6}
\end{figure}

Then, we investigate the collision between two spin solitons. We fail to derive exact two-spin soliton solution, since the model is essentially a non-integrable model. We perform numerical simulations to investigate the collision behavior, based on the integrating-factor method and the fourth-order Runge-Kutta method \cite{Yang}. Since spin soliton is localized in space, the two-spin solitons solution can be seen as approximately a linear superposition of two single spin solitons, if they are far apart from each other. Therefore, we can simulate the collision process from the initial conditions given by exact spin soliton solutions. As an example, we demonstrate the collision between two B-B-DSSs. The initial conditions are chosen as
$\Psi_{+}= c_{2} \sqrt{\frac{g-v_{1}^2}{g}} \textmd{sech}[\sqrt{g-v_{1}^2}(z-z_{10})] e^{i[ v_{1} (z-z_{10})+\Delta\phi]} + c_{1} \sqrt{\frac{g-v_{2}^2}{g}} \textmd{sech}[\sqrt{g-v_{2}^2}(z-z_{20})] e^{i v_{2} (z-z_{20})}$,
$\Psi_{0}= c_{1} \sqrt{\frac{g-v_{1}^2}{g}} \textmd{sech}[\sqrt{g-v_{1}^2}(z-z_{10})] e^{i[ v_{1} (z-z_{10})+\Delta\phi]} + c_{2} \sqrt{\frac{g-v_{2}^2}{g}} \textmd{sech}[\sqrt{g-v_{2}^2}(z-z_{20})] e^{i v_{2} (z-z_{20})}$,
$\Psi_{-}= \frac{(v_{1}+i\sqrt{g-v_{1}^2})(v_{2}+i\sqrt{gc^2-v_{2}^2})}{g^2}(v_{1}-i\sqrt{g-v_{1}^2}\textmd{tanh}[\sqrt{g-v_{1}^2}(z-v_{1}t-z_{10})])
(v_{2}-i\sqrt{g-v_{2}^2}\textmd{tanh}[\sqrt{g-v_{2}^2}(z-v_{2}t-z_{20})])$. We can investigate the collisions between spin soltions with different relative velocities through varying $v_1$ and $v_2$. The parameter $\Delta\phi$  describe the relative phase between bright solitons, the parameters $z_{10}$ and $z_{20}$ are induced to control the initial positions of two spin solitons.

When the soliton velocity is much smaller than the maximum moving speed, the spin density distribution profile varies after the collision. Namely, the solitons in each component admit varied density profiles. For example, we show one case with $v_1=-v_2=0.1$ and $\Delta\phi=\frac{\pi}{2}$ in  Fig.~\ref{Fig5} (a1-a3). Especially, the solitons can admit  weak amplitude and location oscillations after the collision. As far as we know, this behavior is absent for integrable systems \cite{Lakshman,Kanna,Lingdnls,Feng,DSWang,Ohmi,Ieda,Li,WZhang,Uchiyama,Kivshar,Ueda}, due to that integrable systems possesses infinite conservation laws. This character partly means that the model discussed above is indeed a non-integrable model. Interestingly, an elastic collision process can emerge with some special relative phases. We show one case with  $v_1=-v_2=0.1$ and $\Delta\phi=0$ in  Fig.~\ref{Fig5} (b1-b3). It is seen that the two spin solitons firstly approach each other, then bounce back.  Moreover, the relative velocity also affects the spin density redistribution behavior. Keeping identical relative phase $\Delta\phi=0$ with Fig.~\ref{Fig5} (b1-b3), we choose $v_1=-v_2=0.4$  to show the spin density variations in Fig.~\ref{Fig5} (c1-c3).  One can see that two initial spin solitons with distinctive profiles changes to be two spin solitons with almost identical profiles after the collision.   The weak amplitude and location oscillations also appear.

When the soliton velocities are close to the maximum speed, the collision between spin solitons becomes classically elastic. The behavior is similar with the ones in integrable systems \cite{Lakshman,Kanna,Zhao3}.   We show one case with $v_{1}=-v_{2}=0.9$ and $\Delta\phi=0$ (the maximum speed is $\sqrt{g}=1$) in Fig.~\ref{Fig6} (a1-a3). The two solitons' profiles do not vary after their collision and the interference patterns emerge during the interacting process. The relative phase does not affect the elastic collision properties, but vary the interference pattern (see Fig.~\ref{Fig6} (b1-b3)).  This is similar with the interference between dark-bright solitons \cite{Qin2}.

The numerical simulations indicate that the collision between spin solitons can be both elastic and inelastic, which depends on the relative phase and relative velocity between them.  The spin density redistribution happens during the collision process. The non-integrable properties of the systems can make spin soliton experience weak amplitude and location oscillations after the collision.

\section{conclusion}
We present four families of exact analytical spin soliton solutions for non-integrable models, which further enrich the nonlinear excitations greatly in the BEC systems described by non-integrable models.  Numerical simulation results indicate that it is possible to excite these spin solitons in experiments, with the development of experimental techniques \cite{Bersano}. Spin density redistribution happens during the collision between spin solitons, which depends on the relative phase and velocity. They have great potential applications to study the negative inertial mass of solitons \cite{zhaoliu,Qu,PNAS}, the dynamics of soliton-impurity systems \cite{PNAS,impu1,impu2,impu3}, and the spin dynamics in BEC systems \cite{spinwave,swave,swave2}.

\section*{Acknowledgments}
This work is supported by National Natural Science Foundation of
China (Contact No. 11775176), China Scholarship Council, Major Basic Research Program of Natural Science of Shaanxi Province (Grant No. 2018KJXX-094), The Key Innovative Research Team of Quantum Many-Body Theory and Quantum Control in Shaanxi Province (Grant No. 2017KCT-12), and the Major Basic Research Program of Natural Science of Shaanxi Province (Grant No. 2017ZDJC-32).

\end{document}